\documentclass[%
 preprint,
nofootinbib,
 amsmath,amssymb,
 aps,
 pra,
]{revtex4-2}

\usepackage{graphicx}
\usepackage{dcolumn}
\usepackage{bm}
\usepackage{hyperref}
\usepackage{slashed,color,mathtools,amsthm}
\usepackage{comment}
\usepackage{enumerate}
\usepackage{float}
\usepackage{flafter}
\usepackage{placeins}

\usepackage{caption}
\usepackage{subcaption}

\newcommand{\be}{\begin{equation}}
\newcommand{\ee}{\end{equation}}
\newcommand{\bea}{\begin{eqnarray}}
\newcommand{\eea}{\end{eqnarray}}
\newcommand{\beas}{\begin{eqnarray*}}
\newcommand{\eeas}{\end{eqnarray*}}
\newcommand{\ba}{\begin{array}}
\newcommand{\ea}{\end{array}}

\newcommand{\ket}[1]{\left| #1 \right\rangle}

\newcommand{\bra}[1]{\left\langle #1 \right|}

\begin{document}

\title{Entanglement spectrum and Schmidt mode evolution during two-particle scattering}

\author{Arsam Najafian}
 \email{arsamn@student.ubc.ca}
\author{Mark Van Raamsdonk}
 \email{mav@phas.ubc.ca}
\affiliation{%
 Department of Physics and Astronomy, University of British Columbia,\\
 6224 Agricultural Road, Vancouver, B.C., V6T 1Z1, Canada
}%

\date{\today}
\begin{abstract}
A particle initially in a pure state but interacting with some environment evolves into a discrete ensemble of pure states, the eigenstates of its reduced density operator, with ensemble probabilities given by the corresponding eigenvalues. In this work, we use numerics to present explicit results for the time-dependence of these eigenvalues and eigenstates for simple scattering experiments in one and two dimensions. This provides a time-resolved picture of the scattering process, showing in detail how an initial state described entirely in terms of continuous parameters evolves into a discrete set of possible outcomes, each with an associated probability and time-evolving wavefunction. We find that for scattering of Gaussian wavepackets in one dimension, the late time spectrum is dominated by two large eigenvalues nearly equal to the transmission and reflection probabilities associated with the central value of momentum. The corresponding eigenstates appear as single-peaked reflected or transmitted wavepackets. The remaining smaller eigenvalues, which increase to a maximum during scattering and then decrease to small values, correspond to reflected or transmitted wavepackets with multiple spatially separated parts. In this case and also for two-dimensional scattering, we find that successively smaller eigenvalues correspond to probability distributions with successively more peaks. These multi-peaked states correspond to outcomes of the scattering experiment where a particle initially in a single wavepacket ends up in a superposition of separated wavepackets after scattering.
\end{abstract}

\maketitle


\hypersetup{colorlinks = true}
\let\oldphi\phi 
\let\phi\varphi 
\let\varphi\oldphi

\section{Introduction}

For a spin or other discrete quantum system, there is a very familiar story about what happens to such a system that is initially in a pure state but then is measured or more generally interacts with an environment \cite{JoosZeh1985,Zurek2003,Schlosshauer2007,JoosEtAl2003}. Taking the measuring device and/or environment as part of the full quantum system, the state of the subsystem of interest at any time is captured by its reduced density matrix. This is initially a projector with a single non-zero eigenvalue, but then during the interaction with the measuring apparatus and/or environment evolves to a more general form with multiple non-zero eigenvalues. In the measurement scenario, these eigenvalues represent the probabilities of the various measurement outcomes and the corresponding eigenvectors represent the pure states corresponding to these outcomes. These are the possibilities for what the wavefunction would collapse to in the Copenhagen interpretation.

For a continuous quantum system such as a free particle, we can again consider the situation where the system is initially in a pure state and then interacts with a measurement apparatus or environment. Again, we can describe the state of the system at any time by a reduced density matrix, but this is now a more complicated object. For a particle, the density matrix can be described by a function  $\rho(\vec{x},\vec{y},t) = \langle \vec x | \rho(t) | \vec y \rangle$. We can still consider the eigenvalue problem
\begin{equation}
\label{eigenvalue}
\int d^d y \rho(\vec{x},\vec{y},t) \psi(\vec{y}) = \lambda \psi(\vec{x})
\end{equation}
but the behaviour of the eigenvalues and eigenvectors of this operator as a function of time during a measurement process or more general interaction with an environment is less familiar.

The density operator is a self-adjoint, positive, trace-one operator. Any self-adjoint trace-class\footnote{Trace-class means having finite trace norm ${\rm Tr}(\sqrt{{\cal O}{\cal O}^\dagger})$. For a density operator ${\rm Tr}(\sqrt{\rho \rho^\dagger}) = {\rm Tr}(\rho) = 1$.} operator on a separable Hilbert space is compact\footnote{An operator is compact if the image under $A$ of any bounded set in ${\cal H}$ has compact closure in ${\cal H}$.} and any compact self-adjoint operator has a discrete spectrum. For our density operator, this is known as the entanglement spectrum \cite{LiHaldane2008}. 

Since all the eigenvalues must be positive and sum to one, we either have a finite number of non-zero eigenvalues or an infinite number with an accumulation point at 0. We expect the latter to be the generic situation, unless we have a special setup such as an environment with a finite dimensional Hilbert space.\footnote{In this case, the Schmidt decomposition of the joint state shows that the number of non-zero eigenvalues of the reduced density operator cannot exceed the dimension of the environment Hilbert space \cite{NielsenChuang2010}. It is somewhat remarkable that the state of a single particle carries the information about whether the environment Hilbert space is finite dimensional or infinite dimensional.} With an infinite number of non-zero eigenvalues, we can either have no zero eigenvalues, a finite number of zero eigenvalues, or an infinite number. For a generic state in a generic interacting system, we expect no zero eigenvalues. 

\begin{figure}[!htbp]
    \centering
\includegraphics[width=  \linewidth]{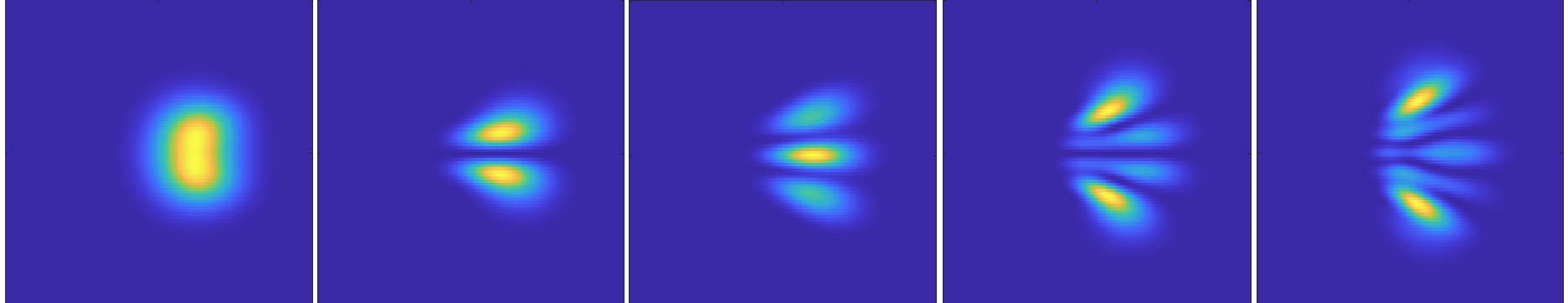}
    \caption{Probability distributions for eigenstates corresponding to the five largest eigenvalues of the density operator for a single particle shortly after scattering from another particle via a repulsive short-range interaction. The particle was initially in a pure state Gaussian wavepacket travelling to the right. These represent a discrete set of possible outcomes of the scattering experiment.}
    \label{fig:FiveGuys}
\end{figure}

During the evolution of the full system, the spectrum of eigenvalues $\{p_n\}$ and the corresponding set of eigenvectors of the density operator for our subsystem will change with time. We expect that generically without symmetries, the spectrum of non-zero eigenvalues is non-degenerate, and further, that the eigenvalues generically do not cross during time evolution as the set of states with degenerate eigenvalues has codimension larger than one in the full space of states. Letting $p_n(t)$ be the $n$th largest eigenvalue as a function of time, we then have an unambiguous corresponding eigenstate $\psi^{sys}_n(t)$ for each non-zero eigenvalue. We can write the evolution of the state of the full system (via the Schmidt decomposition) via a discrete but typically infinite sum
\[
|\Psi(t) \rangle = \sum_n \sqrt{p_n(t)} |\psi^{sys}_n(t) \rangle \otimes |\psi^{env}_n(t) \rangle \; .
\]
It is fascinating that we have the emergence of discreteness here in a physical problem whose description involves only continuous parameters: an initially pure state of a particle evolves into a discrete ensemble of pure states $\psi^{sys}_n$ with probabilities $p_n$. Thus, in the measurement context, we always have a discrete set of outcomes and probabilities similar to the finite-dimensional case.

In this note, we would like to understand explicitly the evolution of eigenvalues $p_n(t)$ and the eigenvectors $\psi^{sys}_n(t)$ for the density matrix of a particle that is initially in a pure state wavepacket but then interacts with an environment consisting of a single other particle. We take the initial state of the system to be a product state where the two particles are described by separated Gaussian wavepackets approaching each other. This is a textbook scattering problem, but as far as we are aware, the evolution of the single-particle density operator eigenvalues and eigenstates has not been previously described (see below for related work). In the typical treatment, we calculate transmission and reflection probabilities in 1D or probabilities for scattering to various angles in higher dimensions. But as we have described, there is a more detailed time-resolved picture where each particle's initial pure state evolves into a discrete ensemble $\{(p_n(t),|\psi_{n}(t)\rangle )\}$. Our goal is to explicitly plot the evolution of the probabilities $p_n(t)$ and the corresponding probability densities $|\psi(\vec{x},t)|^2$ for the largest eigenvalues in a variety of scattering scenarios. 

Even in cases where the scattering problem can be solved analytically, the problem of finding eigenvalues and eigenvectors of the reduced density operator is one of solving the eigenvalue problem
(\ref{eigenvalue}) to find the spectrum of an operator on an infinite dimensional Hilbert space. This cannot be treated analytically. We approach it with a numerical scheme where we introduce periodic boundary conditions, work in a discrete momentum basis, and truncate the Hilbert space to some finite range $[-n_{max},n_{max}]$ for each discrete momentum variable. We evolve the system only for times where the wavefunctions remain well-localized away from the artificial boundaries, and choose $n_{max}$ large enough so that the results do not change significantly as this is increased. Taking advantage of the block-diagonal nature of the Hamiltonian due to overall momentum conservation, we are able to work with Hilbert space dimensions of up to $10^6$. 

\begin{figure}[!htbp]
    \centering
\includegraphics[width= 0.7 \linewidth]{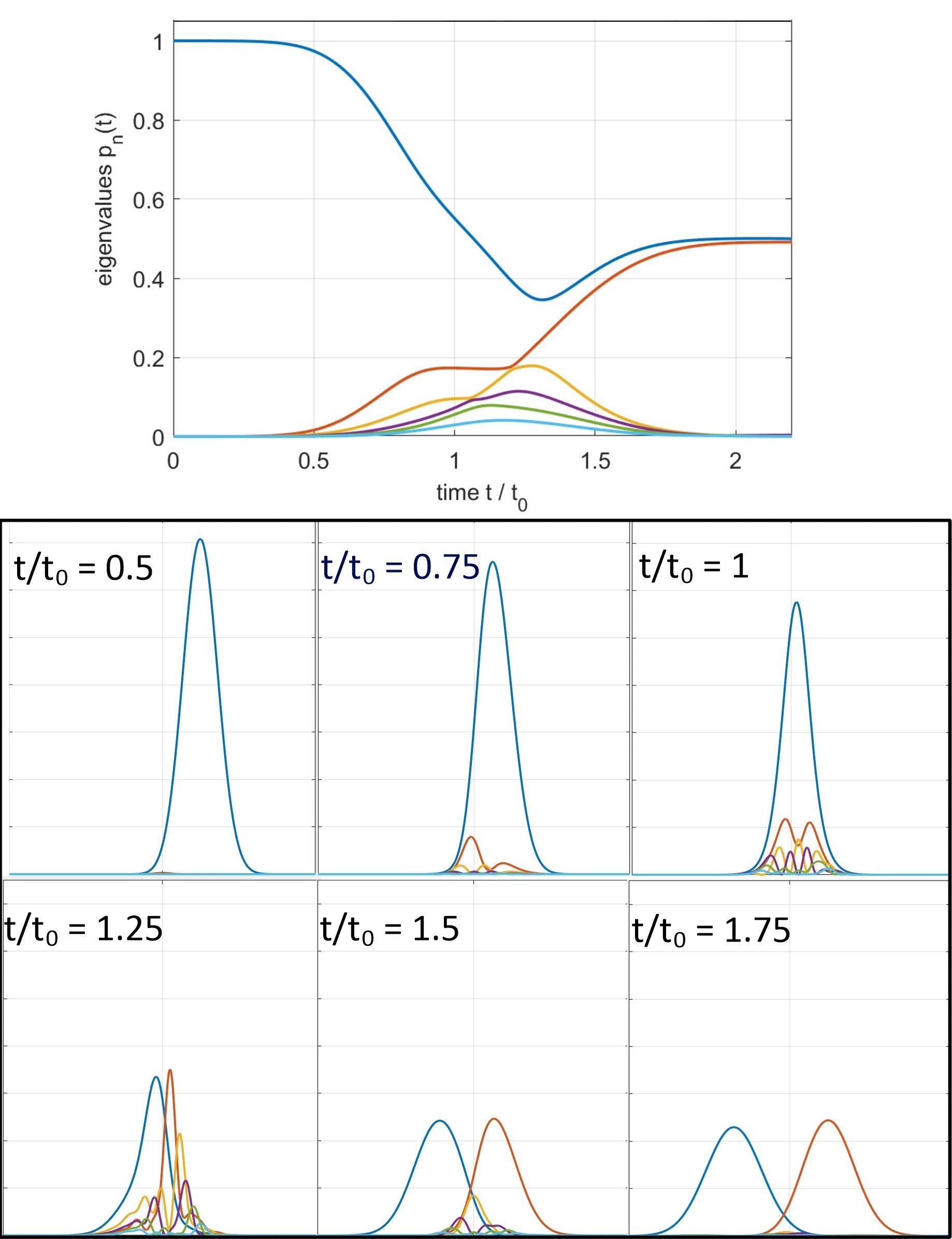}
    \caption{Evolution of the $p_i(t) |\psi_i(x,t)|^2$ for the eigenvectors $\psi_i(x,t)$ of the single particle reduced density matrix for a particle scattering symmetrically off another particle via a short range repulsive Gaussian interaction potential. The interaction strength corresponds to transmission probability $T \approx 0.5$.}
    \label{fig:G356combo}
\end{figure}

We find a variety of interesting qualitative results. In all cases, we find that starting with the initial eigenvalue spectrum $\{1,\vec{0}\}$, the spectrum evolves to ``grow'' more non-zero eigenvalues that eventually approach some limiting values $\{p_i^{out}\}$. For the 1D scattering examples, we find that the asymptotic values of $p_1$ and $p_2$ are very close to the standard reflection and transmission probabilities (with $p_1$ close to the larger of these) associated with the central value of initial momentum. The remaining eigenvalues have the behaviour that they grow to some maximum value during scattering but then decay again to some very small values (see Figures \ref{fig:G356combo} and \ref{fig:Logplot}). The eigenstates corresponding to $p_1$ and $p_2$ have probability distributions that look like Gaussian wavepackets with a single peak, while the eigenstates for the smaller eigenvalues each appear to be either transmitted or reflected waves (rather than superpositions). For either the reflected set or the transmitted set, eigenstates with successively lower probabilities are found to have probability distributions with successively more peaks (Figure \ref{fig:Lowmodes}). For the 2D scattering, we find that the eigenvalues evolve monotonically to their asymptotic values, and that again, the eigenstates for successively lower probabilities have successively more peaks in their probability distribution. An example of this is shown in Figure \ref{fig:FiveGuys}. Videos of the eigenstate evolution for 1D and 2D scattering may be found here \url{https://youtu.be/ZNBMecZCz5A}.

Our results provide a detailed picture of the basic interactions experienced by a particle in a dilute environment where the interactions are typically a sequence of two-particle interactions. The full evolution of a particle's density operator eigenvalues and associated eigenstates in such an environment can be understood as a sequence of discrete branching events where each eigenstate of the density operator after a given scattering event evolves into a new ensemble after the next scattering event. Each of these individual branchings should be well-modelled by the type of two-particle interactions that we study, so our results and similar future studies should help provide a realistic and detailed picture of what happens to a particle in an environment consisting of a dilute gas of other particles.

As one further application, we employ our techniques to analyze a microscopic model of a measurement where a 1D particle initially in a spatial superposition state moving to the left (the system) interacts with a localized particle (the probe) moving to the right. We find that the system particle ends up in a mixed state with only two significant eigenvalues, one corresponding to a situation with no scattering where the wavefunction continues on as a spatial superposition moving to the left and the other corresponding to a situation with scattering where the system particle's wavefunction has collapsed to a single localized Gaussian moving to the right. In the latter case, the probe particle's outgoing wavefunction carries the information about the system particle's initial superposition state.

The details of the scattering problem and our numerical approach are described below in section 2, and our results are described in detail in section 3. In section 4, we provide an analytical demonstration showing why in our one-dimensional scattering analyses only two significant eigenvalues are observed in the final state. We end with a brief discussion in section 5.

\subsection*{Relation to previous work}

Our present investigation is related to previous works that have studied the evolution of entanglement in scattering experiments \cite{law2004entanglement, SchmuserJanzing2006,HarshmanHutton2008,Weder2011,Weder2013,PeschanskiSeki2016}. These analyses characterize the evolution and/or final entangled state in terms of the purity or von Neumann entropy of a single-particle reduced density matrix. These measures provide a coarse-grained description of the density matrix eigenvalue evolution in terms of single numbers ($\sum_i p_i^2$ for the purity or $-\sum_i p_i \log p_i$ for the von Neumann entropy) calculated from the spectrum of the density operator. 
However, to our knowledge, our more detailed combination of time-resolved entanglement spectrum and real-space eigenfunctions for a continuous-variable scattering problem has not been presented before. Our work is also related to many other works studying microscopic models of decoherence. The full set of references would be too numerous to list, but we note that  \cite{schulman2004evolution} considered the evolution of the spread of the wavepacket during scattering with environmental particles. 

\section{Basic setup}

For our investigations, we consider a ``system'' particle of mass $m_1$ that interacts with an environmental particle of mass $m_2$ via a potential that depends only on the separation between the particles. The total Hamiltonian is given by \eqref{ho}.
\begin{equation}\label{ho}
    H=\frac{\vec{p}_1^2}{2m_1}+\frac{\vec{p}_2^2}{2m_2} + V(|\vec{x}_1-\vec{x}_2|)
\end{equation}
We will consider some initial product state 
\[
\Psi(\vec{x}_1, \vec{x}_2) = \psi_1(\vec{x}_1) \psi_2(\vec{x}_2)
\]
where $\psi_1$ and $\psi_2$ are taken to be Gaussian wavepackets initially approaching each other. 

Explicitly, we consider initial single-particle states $|\psi_i\rangle = |\vec{k}^0_i, \Delta_i, \vec x^0_i \rangle$ where
\[
\langle \vec{x} |\vec{k}, \Delta, \vec x^0 \rangle  = {\cal N} \int d^d \vec{k} e^{i \vec{k} \cdot (\vec{x}- \vec{x}_i)} e^{-(\vec{k} - \vec{k}^0)^2 \over (2 \Delta)^2} = \hat{\cal N} e^{- \frac{\sigma}{2} (\vec{x}- \vec{x}^0)^2} e^{i \vec{k}^0 \cdot (\vec{x}- \vec{x}^0)}
\]
and ${\cal N}$ and $\hat{\cal N}$ are the normalization constants. We can formally solve for the evolution of the wavefunction by introducing center-of-mass and relative coordinates so that the Hamiltonian splits as $H = H_{CM} + H_{rel}$. The energy eigenstates for the center of mass Hamiltonian are total momentum eigenstates $|\vec{K}\rangle$, while the energy eigenstates for the relative Hamiltonian will in general be a discrete set of bound states $|n \rangle$ with $E_n \le 0$ and scattering states $|\vec{k}\rangle$ that we can label by the incoming momentum $\vec{k}$. Expanding the initial state in terms of the energy eigenstates $|\vec{K},n\rangle$ and $|\vec{K},\vec{k}\rangle$, we can time evolve as usual by multiplying the eigenstate coefficients by phases $e^{-i E t / \hbar}$. Finally, the time-dependent density operator for the first particle can be computed as
\[
\rho_1(\vec{x}_1,\vec{x}_1',t) = \int d^d x_2 \Psi(\vec{x}_1,\vec{x}_2 ,t) \Psi(\vec{x}_1',\vec{x}_2 ,t)^*
\]
We are interested in the time evolution of the eigenvalues and eigenvectors for this density operator, as defined by (\ref{eigenvalue}).

\subsection{Numerical setup}

Even for potentials where we can obtain the bound state and scattering state wavefunctions analytically, the problem of solving the infinite-dimensional eigenvalue problem (\ref{eigenvalue}) to find the spectrum  appears intractable analytically, so we will resort to numerical methods. In order to facilitate these, we regulate the problem by setting it in a finite volume box with side lengths of $L$ and periodic boundary conditions. 

In this case, we have discrete momentum eigenstates $|\vec{n} \rangle$ for each particle with momentum $2 \pi \hbar \vec{n}/L$ and position space wavefunction
\[
\bra{\vec{x}} \vec{n} \rangle = {1 \over L^{d \over 2}} e^{2 \pi i \vec{n} \cdot \vec{x} / L} \; .
\]
In the product basis $|\vec{n}_1, \vec{n}_2 \rangle$, the Hamiltonian becomes
\[
\bra{\vec{n}_1,\vec{n}_2}  H  \ket{\vec{n}_1',\vec{n}_2'} = \left( { 2 \pi^2 \hbar^2  \vec{n}_1^2 \over  m_1 L^2} +{2 \pi^2 \hbar^2 \vec{n}_2^2 \over m_2 L^2}\right) \delta_{\vec{n}_1,\vec{n}_1'} \delta_{\vec{n}_2,\vec{n}_2'} + \hat{V}(\vec{n}_1' - \vec{n}_1) \delta_{\vec{n}_1 + \vec{n}_2, \vec{n}_1' + \vec{n}_2'} 
\]
Where the last delta function comes from momentum conservation and the remaining part of the momentum space potential is 
\begin{equation}\label{Vterms}
    \hat{V}(\Delta \vec{n}) = {1 \over L^d} \int d^d x V(|\vec{x}|) e^{2 \pi i \Delta \vec{n} \cdot \vec{x} \over L}
\end{equation}

In the finite volume setup, we take the initial state for each particle to be a wavepacket defined in terms of the momentum eigenstates $|\vec{n} \rangle$ by
\[
|N_c, X_c, \sigma \rangle = {\cal N}\sum_{n \in \mathbb{Z}}e^\frac{-(\vec{N}_c-\vec{n})^2}{(2\sigma)^2} e^{-\frac{2 \pi i \vec{X_c} \cdot \vec{n}}{L}}\ket{\vec{n}}
\]
where $\vec{N}_c$ and $\sigma$ determine the central value and uncertainty in momentum, $X_c$ is the initial position, and ${\cal N}$ is the normalization. 

In all of our analyses below, we consider the symmetrical situation with particles of equal mass ($m_1 = m_2$) and equal $\sigma$ approach each other with initial positions at $\vec{X} = (\pm x_c,\vec{0})$ and initial central momenta $\vec{N}_c = (\mp n_c, \vec{0})$. 

The initial uncertainty in physical momentum is
\[
\Delta p = {2 \pi \hbar \sigma \over L}
\]
while the initial uncertainty in position is
\[
\Delta x = {L \over 4 \pi \sigma} \; .
\]
We will choose $\sigma > 1$ so that the wavepacket widths are significantly smaller than the circle size and choose the initial separation to be large enough so that the wavepackets do not overlap significantly in the initial state. 

The central particle velocity is
\[
v = {2 \pi \hbar n_c \over L M} \; ,
\]
so the time when the wavepackets coincide when interactions are set to zero is
\[
t_0 = {x_c L M\over 2 \pi \hbar n_c} \; .
\]
We would like the spreading of the wavepacket during this collision time to be small compared to the initial separation (which is of the order of the initial wavepacket size). This requires $
t_0 \Delta v \ll x_c$, which gives
\[
\sigma \ll n_c \; .
\]
In the results presented below, we ensure that the states are well-localized to the interval $[-L/2,L/2]$ during the full evolution so that the results should be the same as for scattering on a line. We test this by verifying that the results remain the same after adjusting $L \to 2L$, keeping the physical momentum parameters $n_c/L$ and $\sigma/L$ fixed.

For our calculations, we work with a momentum cutoff, keeping only momentum components $-n_{max} \le n_i \le n_{max}$. We take $n_{max}$ large enough so that further increases in $n_{max}$ do not substantially change the results.

Since the interaction depends only on the relative coordinate, total momentum is conserved, and the Hamiltonian decomposes into blocks corresponding to specific total momenta. These blocks are significantly smaller than the full matrix, and can be diagonalized independently. The initial two-particle wave packet is expanded in the eigenbasis for each block, and evolved by multiplying the components by phases $e^{-i E_n t/\hbar}$. The reduced density operator for one particle is obtained directly from this time-evolved  state. Diagonalization of the reduced density matrix yields the time-dependent entanglement spectrum and the corresponding single-particle eigenmodes, which are transformed back to real space for visualization. 

This approach is numerically efficient and gives direct access to both eigenvalues and eigenfunctions of the reduced density operator throughout the scattering process.

\section{Results}

In this section, we present the results for the evolution of reduced density operator eigenvalues (the entanglement spectrum) and eigenstates for particle scattering with short-range interactions in one and two dimensions. We always consider symmetrical scattering events where the particles have equal mass and approach each other in equivalent initial Gaussian wavepackets.

\subsection{One dimension}

First, we consider one-dimensional scattering. We choose a short-range potential
\begin{equation}
\label{Gpot}
V(r) = {A \over \sqrt{2 \pi} w} e^{-{x^2 \over 2 w^2}} 
\end{equation}
with range $w$, which approaches the delta function potential
\[
V(r) = A \delta(x) \; .
\]
in the limit where $w \to 0$.

\subsubsection*{Delta function limit}

In the limit $w \to 0$ where we have a delta function potential, standard 1D scattering calculations show that the transmission probability is (see e.g. \cite{GriffithsSchroeter2018})
\[
T = {1 \over 1 + \left({\mu A \over \hbar  p } \right)^2}
\]
where $p$ is the physical momentum of each initial particle and $\mu$ is the reduced mass.

\begin{figure}[!htbp]
    \centering
\includegraphics[width=\linewidth]{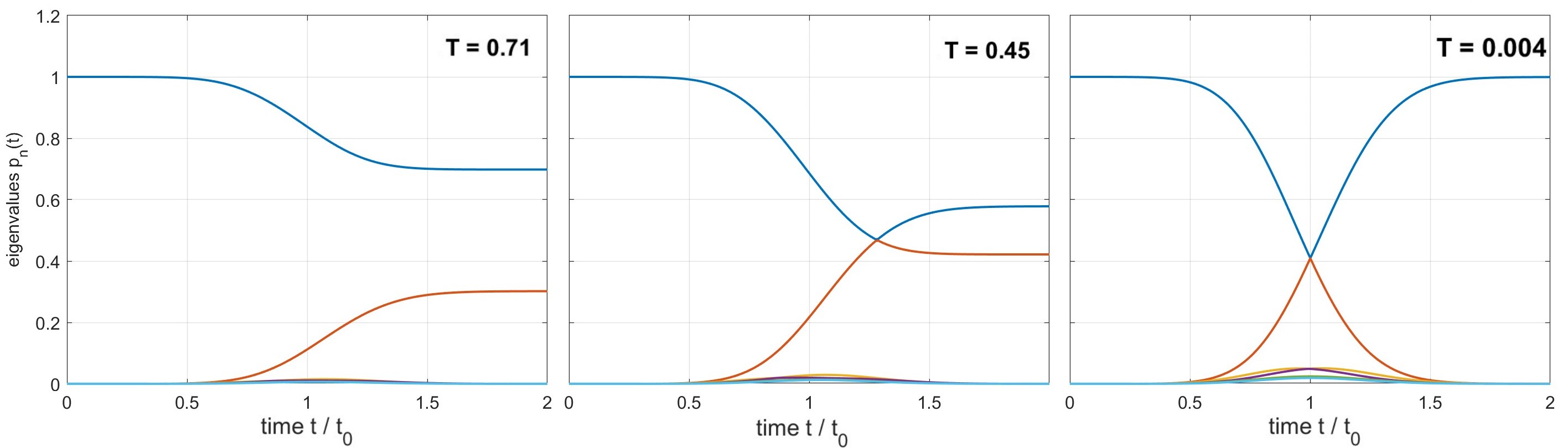}
    \caption{Evolution of the eigenvalues of the reduced density operator for a particle scattering symmetrically off another particle via a delta function interaction, for three different values of the interaction strength, parameterized by the transmission probability $T$. Here $t_0$ is the time at which the initial wavepackets would coincide in the absence of interactions.}
    \label{fig:Deltacombo}
\end{figure}

In Figure \ref{fig:Deltacombo}, we show the evolution of the eigenvalues of the single-particle density operator for three choices of the interaction strength $A$, corresponding to transmission probabilities $T=0.71$, $T=0.45$, and $T=0.004$ (using the central value of momentum). We have chosen initial positions $X_c = \pm L/4$, $\sigma/n_c = 0.15$, $n_{max} = 151$, and $\Delta x = 0.05 L$.

We see that in each case, the eigenvalues evolve from the initial distribution $\vec{p} = (1,\vec{0})$ corresponding to the pure state to a distribution that becomes constant at late times, after the wavepackets for the two particles no longer overlap significantly. The largest two eigenvalues at late times are almost exactly equal to the transmission and reflection probabilities that can be calculated analytically. For $T < 0.5$, we see that the largest two eigenvalues cross near the time when the packet centers coincide. As we mentioned in the introduction, such crossing is not expected for a generic path through the space of density operators, but in this case, the path is non-generic since the Hamiltonian for the system includes translation symmetry, reflection symmetry, and symmetry under particle exchange.

\begin{figure}[!htbp]
    \centering
\includegraphics[width= 0.7 \linewidth]{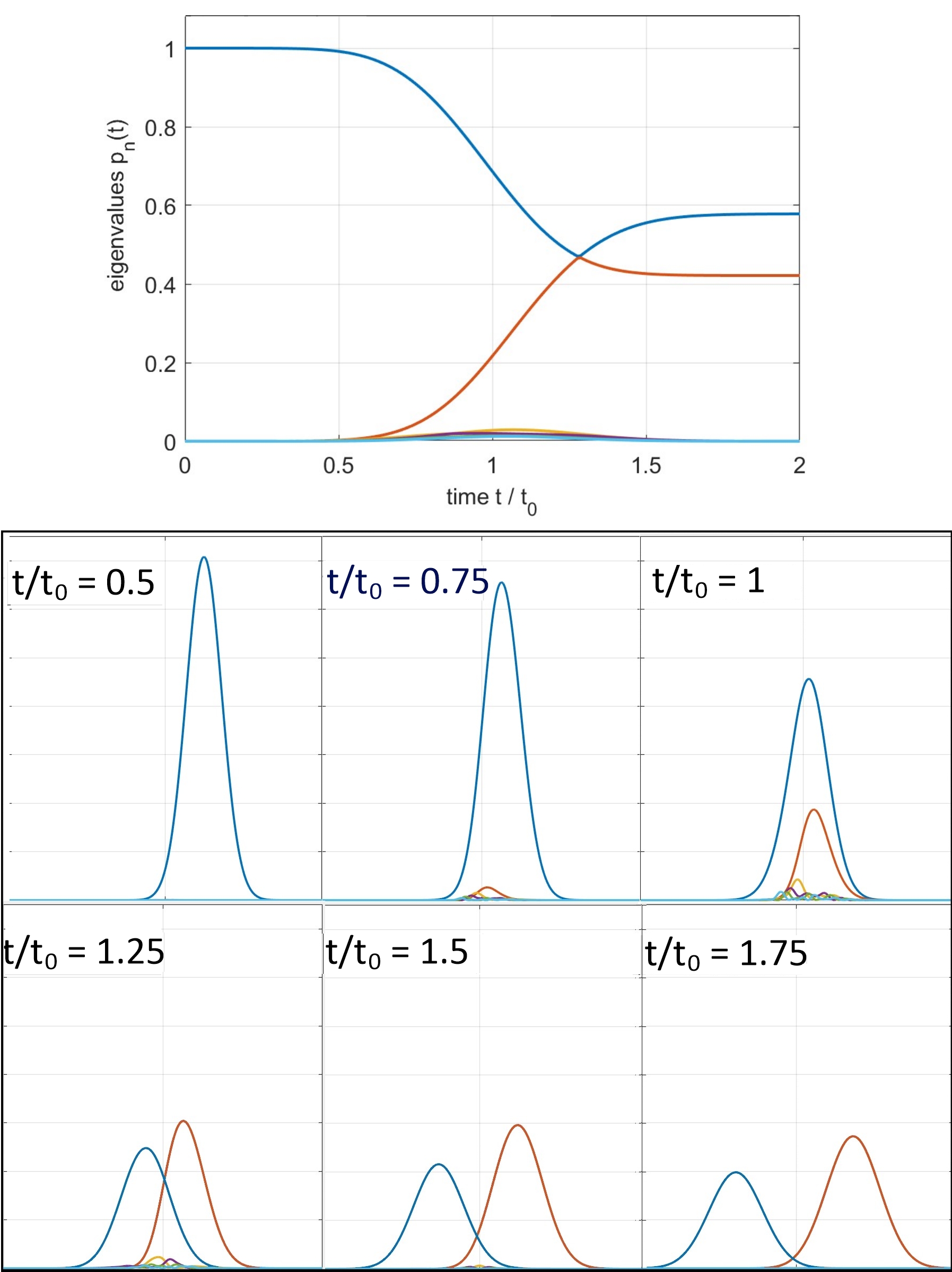}
    \caption{Evolution of the $p_i(t) |\psi_i(x,t)|^2$ for the eigenvectors $\psi_i(x,t)$ of the single particle reduced density matrix for a particle scattering symmetrically off another particle via a repulsive delta function interaction. The interaction strength corresponds to transmission probability $T \approx 0.45$. At late times, the two largest eigenvalues are very close to the reflection and transmission probabilities, and the corresponding eigenvectors have probability distributions as expected for reflected and transmitted wavepackets.}
    \label{fig:D140combo}
\end{figure}

It is illuminating to plot the probability distributions for the eigenvectors of the density operator. In Figure \ref{fig:D140combo}, we plot $p_i(t) |\psi_i(x,t)|^2$ for the eigenvectors $\psi_i(x,t)$ of the single-particle density operator. The normalization by $p_i(t)$ is chosen so that the sum of all the probability distributions integrates up to 1. We see that the initial wavepacket (the eigenfunction for the eigenvalue $p=1$) evolves continuously to become the eigenfunction for the eigenvalue that becomes approximately equal to the transmission probability at late times. The probability density for this eigenfunction appears as a wavepacket moving continuously to the left. The function $p_i(t) |\psi_i(x,t)|^2$ for the eigenvalue that ends up as the largest one, originates near the scattering center and moves to the right to become the reflected wavepacket. 

Apart from the two largest eigenvalues of the density operator, the remaining eigenvalues achieve a maximum value near the time $t_0$ where the two wavepackets would coincide in the absence of interactions and then decrease to become very small at late times. For stronger interactions, the sum of these small eigenvalues becomes larger. We find numerically that for the wavepacket parameters above, the sum of the smaller eigenvalues achieves a maximum of around 0.2 in the limit of large $A$. 

\begin{figure}[!htbp]
    \centering
\includegraphics[width= 0.7 \linewidth]{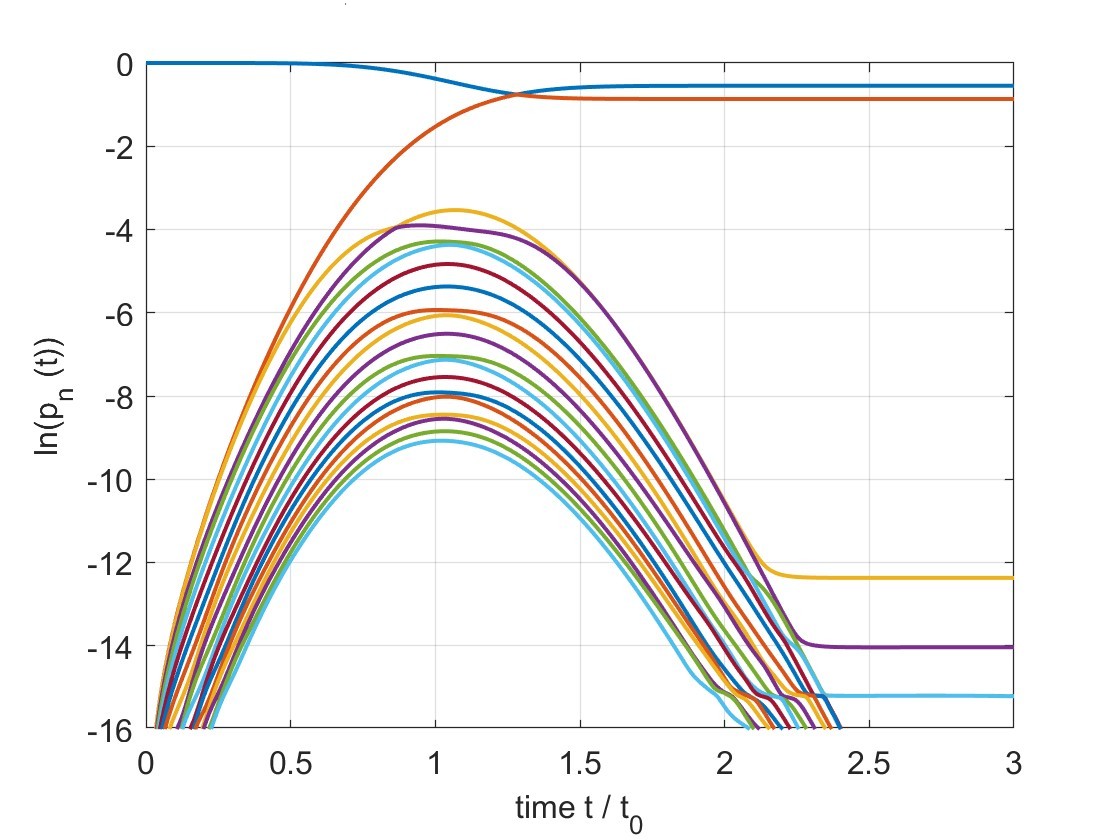}
    \caption{Evolution of the natural logarithm of the largest 20 eigenvalues of the reduced density operator for a particle scattering symmetrically off another particle via a repulsive delta function interaction. Initial wavepackets are taken to be Gaussian. A truncation to Hilbert space dimension 363,609 ($|n_{max}|=301$) was used for this plot.}
    \label{fig:Logplot}
\end{figure}

Figure \ref{fig:Logplot} shows the evolution of the logarithm of the eigenvalues for the scattering parameters above with $T=0.45$. The asymptotic value of the third-largest eigenvalue is $p_3 \approx 10^{-5}$ in this case. The fact that the two largest eigenvalues dominate is related to previous observations \cite{law2004entanglement,SchmuserJanzing2006} that in the limit a hard-wall potential ($A \to \infty$, the final state of the two particles is unentangled).

\begin{figure}[!htbp]
    \centering
\includegraphics[width= 0.8 \linewidth]{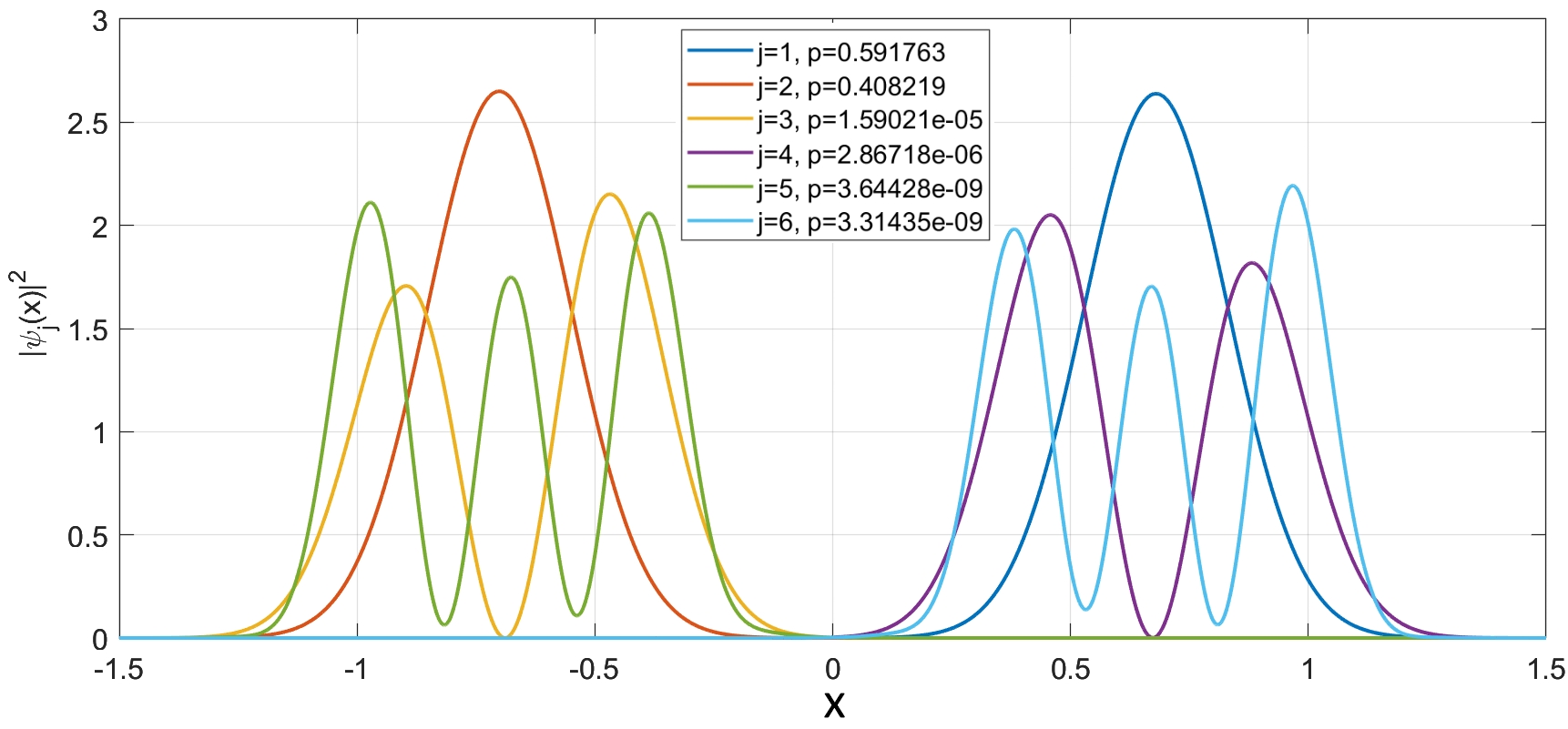}
    \caption{Probability distributions $|\psi_i(x,t)|^2$ for the eigenvectors $\psi_i(x,t)$ of the single particle reduced density matrix at $t=3.75 t_0$ corresponding to the six largest eigenvalues. The eigenstates with successively lower probabilities are observed to have successively more peaks in their probability distribution.}
    \label{fig:Lowmodes}
\end{figure}

It is interesting to look at the eigenstates corresponding to the small eigenvalues, since it is not clear from Figure \ref{fig:D140combo} what happens to these at late times. In Figure \ref{fig:Lowmodes}, we plot the probability distributions for the eigenstates corresponding to the six largest eigenvalues. We find that three of the modes correspond to transmitted wavepackets and three of the modes correspond to reflected wavepackets. In each of these groups, the $k$ largest eigenvalue is observed to have $k$ peaks in its probability distribution, thus the eigenstate has the interpretation of a particle being in a superposition state of $k$ distinct locations, each with an associated uncertainty. We will see a similar phenomenon for the 2D scattering that we study below.

\FloatBarrier

\subsubsection*{Finite width interactions}

We also considered the case with finite width interactions with length scale much smaller than the circle size. For a width $w =  L/ (20 \sqrt{2})$ in the potential (\ref{Gpot}), we choose a value of the interaction strength $A$ for which an analysis of the scattering problem predicts a transmission probability $T = 0.5$.  

Again, for late times, we have only two significant eigenvalues, corresponding to transmitted and reflected wavepackets. However, compared to the delta function interaction potential, the other eigenvalues are much more significant at intermediate times. In the midst of scattering, we have an ensemble of many different possible wavefunctions with significant probabilities. This ensemble evolves to leave only two wavefunctions (a transmitted one and a reflected one) each with probability around 0.5 at late times.

\FloatBarrier

\subsection{Two dimensions}

We have performed a similar analysis in the case of two-dimensional scattering. This case is particularly interesting since the possible outcomes are not at all discrete as the particle can scatter to any angle. Yet the late-time state of the particle is a discrete ensemble of possibilities each with an associated probability and wavefunction. It is interesting to see what these look like.

We have chosen to consider a short-range repulsive potential of the form (\ref{Gpot}).  We take initial parameters $x_c = \pm L/4$, $\sigma/n_c = 0.15$, $n_{max} = 15$, and $\Delta x = 0.05 L$. Note that with $n_{max}=15$ we have 31 allowed values for both $x$ and $y$ momenta for each particle, so the truncated two-particle Hilbert space has dimension $31^4 = 923521$. Despite the full truncated Hamiltonian being of size $923521 \times 923521$, the fact that it splits into blocks of fixed total momenta means that we can still perform the diagonalization and produce eigenvalue/eigenvector plots in a short time on a laptop using Matlab.

\begin{figure}[!htbp]
    \centering
\includegraphics[width=0.6\linewidth]{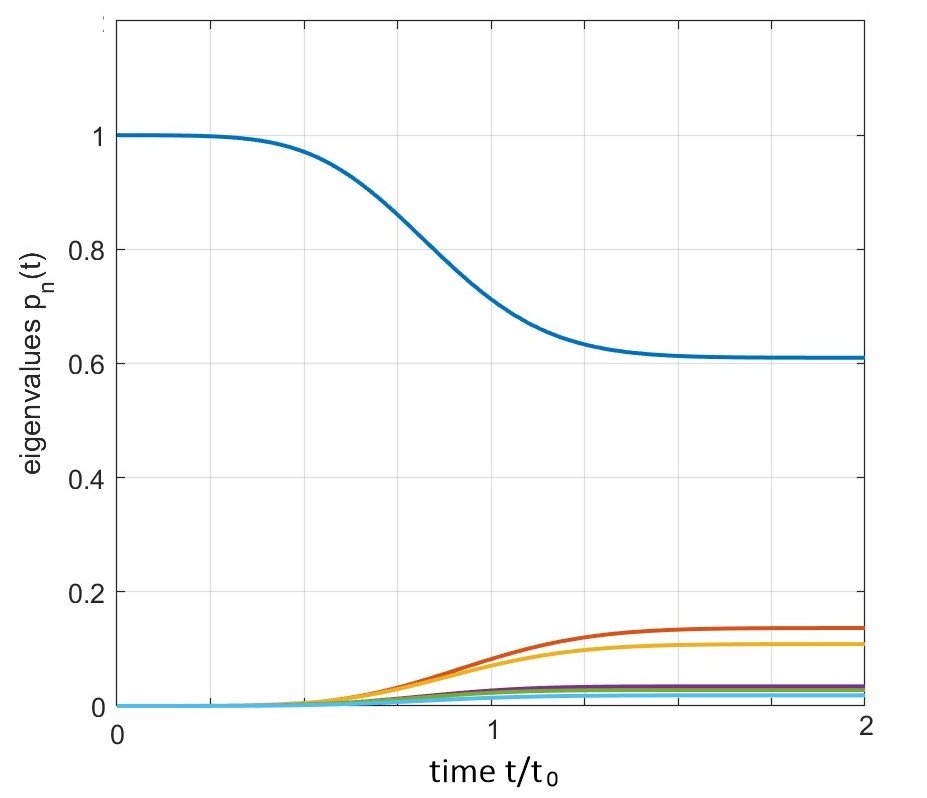}
    \caption{Evolution of the eigenvalues of the reduced density operator for a particle scattering symmetrically off another particle via a short-range repulsive interaction in two dimensions.}
\label{fig:2Deigenvalues}
\end{figure}

The evolution of the eigenvalues is shown in Figure \ref{fig:2Deigenvalues}. We see that the smaller eigenvalues grow from zero and approach constant values at late times when the particles have separated. Unlike the one-dimensional case, we find that this evolution is monotonic.

\begin{figure}[!htbp]
    \centering
\includegraphics[width=0.95\linewidth]{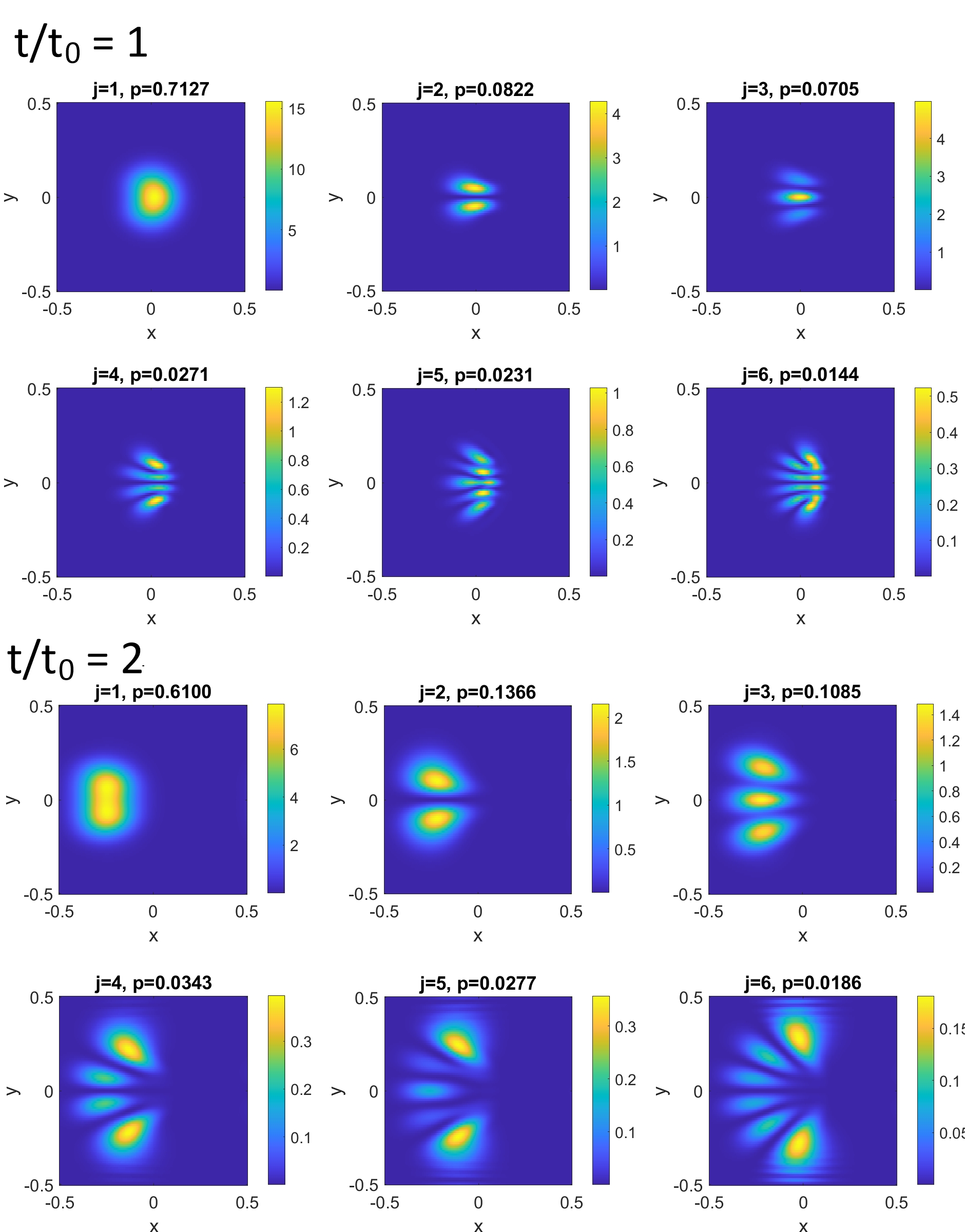}
    \caption{Evolution of probability densities $|\psi_i|^2$ for eigenvectors of the single-particle density operator in 2D scattering with a short-range repulsive potential.}
\label{fig:2Dplots}
\end{figure}

The evolution of the density operator eigenvectors corresponding to the largest six eigenvalues are shown in Figure \ref{fig:2Dplots}. We see that the eigenvector for the $k$th eigenvalue is a wavefunction with $k$ maxima in its probability density at late times, very similar to the phenomenon we observed in one-dimensional scattering. For our particular example, wavefunctions for the smaller eigenvalues are more spread out in the angular direction. 

\subsection{Superposition state scattering}

In the previous examples, we considered initial states consisting of simple Gaussian wavepackets. A natural extension is to consider an initial state where a particle is already in a superposition of spatially separated wavepackets before scattering. This scenario serves as a simple microscopic model for quantum measurement or decoherence, where a particle in a spatial quantum superposition state is probed by another particle. 

We model the interaction by taking the initial state of the system to be a product state where one particle is described by a superposition of two Gaussian wavepackets separated by a large distance, while the second particle is described by a single Gaussian wavepacket approaching from the opposite direction. Explicitly, the initial state for the first particle defined in terms of the momentum eigenstates $|n\rangle$ is modified to
\begin{equation}
    | \psi_{1} \rangle \propto \sum_{n \in \mathbb{Z}} e^{-\frac{(n_{c,1} - n)^2}{(2\sigma)^2}} \left( e^{-\frac{2\pi i X_{1a} n}{L}} + e^{-\frac{2\pi i X_{1b} n}{L}} \right) |n\rangle
\end{equation}
where $X_{1a}$ and $X_{1b}$ are the initial positions of the two peaks. We choose the separation $|X_{1a} - X_{1b}|$ to be large enough such that the interactions of the second particle with the individual peaks in the first particle's wavefunction are
well separated in time. Specifically, we set the simulation domain size to $L=160$, the particle mass to $M=1$, and the wavepacket momentum width to $\sigma=6.0$. The initial positions are chosen as $X_2 = -4.0$ for the incoming particle and $X_{1a} = 4.0, X_{1b} = 44.0$ for the two peaks of the superposition state. 

\begin{figure}[htbp]
    \centering
    \includegraphics[width=0.7\textwidth]{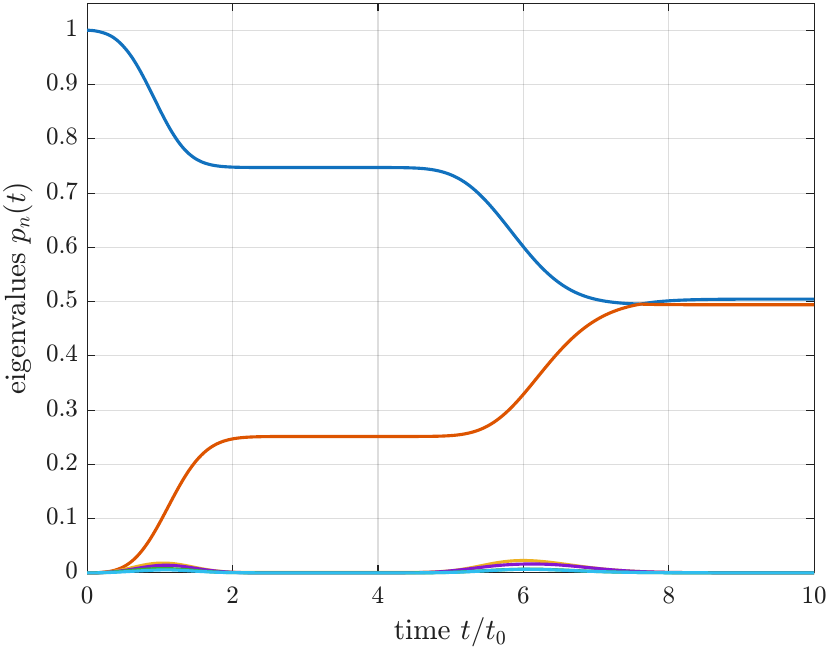}
    \caption{Evolution of the eigenvalues of the reduced density operator for a particle initially in a double-peak superposition state scattering off another particle via a Dirac delta interaction ($n_{c,1} = n_{c,2} = 50$). The two distinct drops in the largest eigenvalue correspond to the two sequential scattering events.}
    \label{fig:double_peak_eigenvalues}
\end{figure}

We consider a short-range repulsive Dirac delta potential $V(r) = A \delta(r)$ and tune the interaction strength $A$ such that a single collision would result in approximately $50\%$ transmission and $50\%$ reflection. We present the results for a symmetric collision where the wavepackets approach each other with equal speeds, corresponding to initial central momenta $n_{c,1} = n_{c,2} = 50$. The evolution of the eigenvalues of the reduced density operator for this setup is shown in Figure \ref{fig:double_peak_eigenvalues}.

\begin{figure}[htbp]
    \centering
    \includegraphics[width=\textwidth]{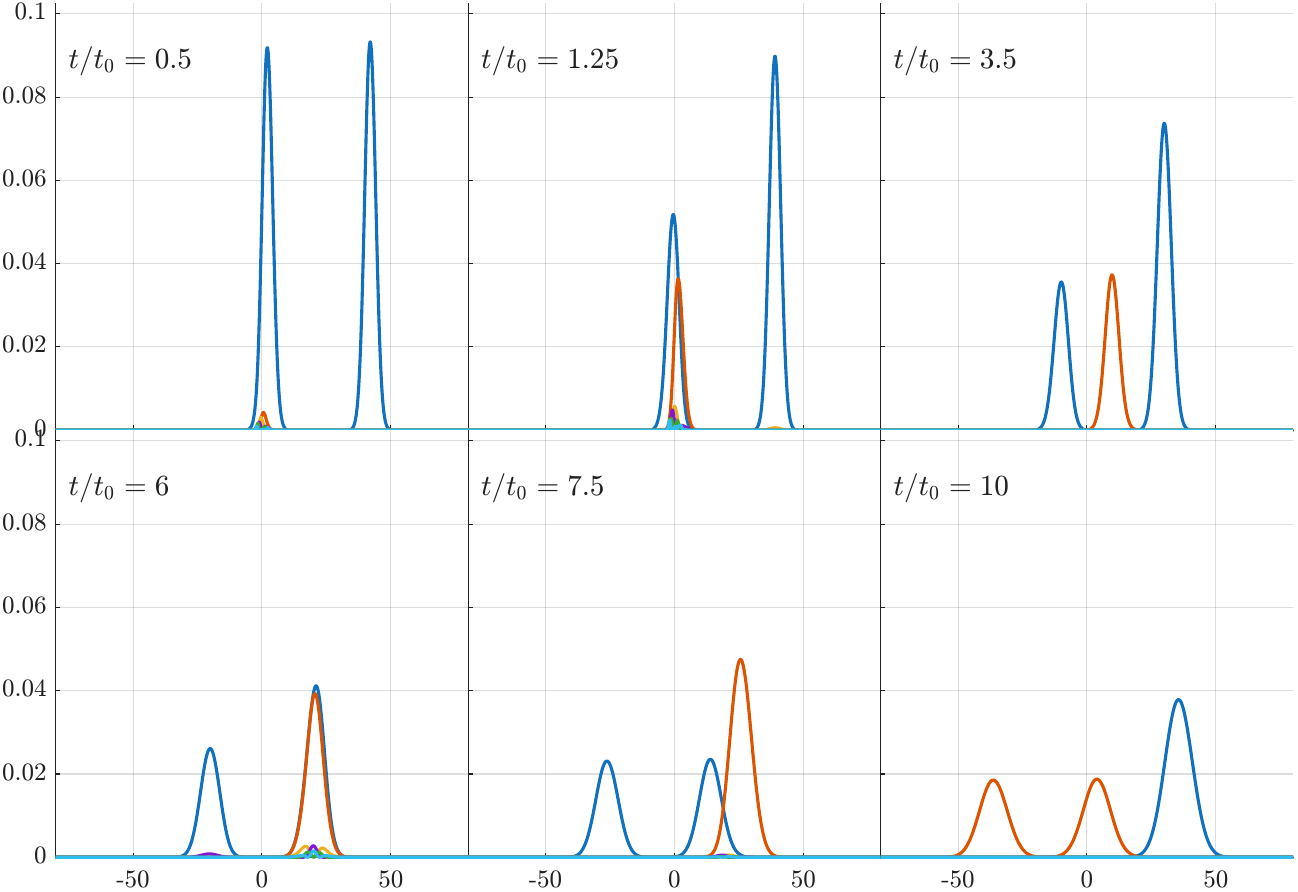}
    \caption{Evolution of $p_i(t)|\psi_i(x,t)|^2$ for the eigenvectors of the reduced density operator of the first particle (initially in a superposition state) during scattering. The time slices capture the state before the first collision ($t/t_0=0.5$), after the first collision ($t/t_0=1.25$), between collisions ($t/t_0=3.5$), during the second collision ($t/t_0=6.0$), and at late times ($t/t_0=10.0$).}
    \label{fig:double_peak_snapshots_p1}
\end{figure}

To visualize the real-space dynamics, we plot the probability distributions for the eigenvectors of the reduced density operator in Figure \ref{fig:double_peak_snapshots_p1}.

Figure \ref{fig:double_peak_eigenvalues} shows that the particle, initially in a pure spatial superposition state, undergoes a two-step scattering process with the second particle: 
\begin{itemize}
    \item Initial: spatial superposition state, moving to the left.
    \item After interaction with first peak: mixed state with $p \approx 0.25$ for a localized wavepacket moving to the right and $p \approx 0.75$ for a spatial superposition state moving to the left with reduced weight in the leading peak.
    \item Final situation: mixed state with $p \approx 0.5$ for a single localized wavepacket moving to the right and $p \approx 0.5$ for an equal amplitude spatial superposition state moving to the left.
\end{itemize}
Thus, although we are dealing with a continuous system whose density operator has an infinite number of eigenvalues, the state resulting from the scattering is largely a mixture of two possibilities, one where no scattering happens and the initial wavefunction remains in a spatial superposition moving to the left, and another where scattering happens and we have a wavefunction that has ``collapsed'' to a single peak moving to the right. This is somewhat surprising: we might have expected that the second particle measuring the first at its two possible locations to correspond to different outcomes with their own probabilities, but what we see is that ``scattering from the first peak'' and ``scattering from the second peak'' add at the level of amplitudes to give a single localized scattered wavepacket with a single associated probability.

\begin{figure}[htbp]
    \centering
    \includegraphics[width=\textwidth]{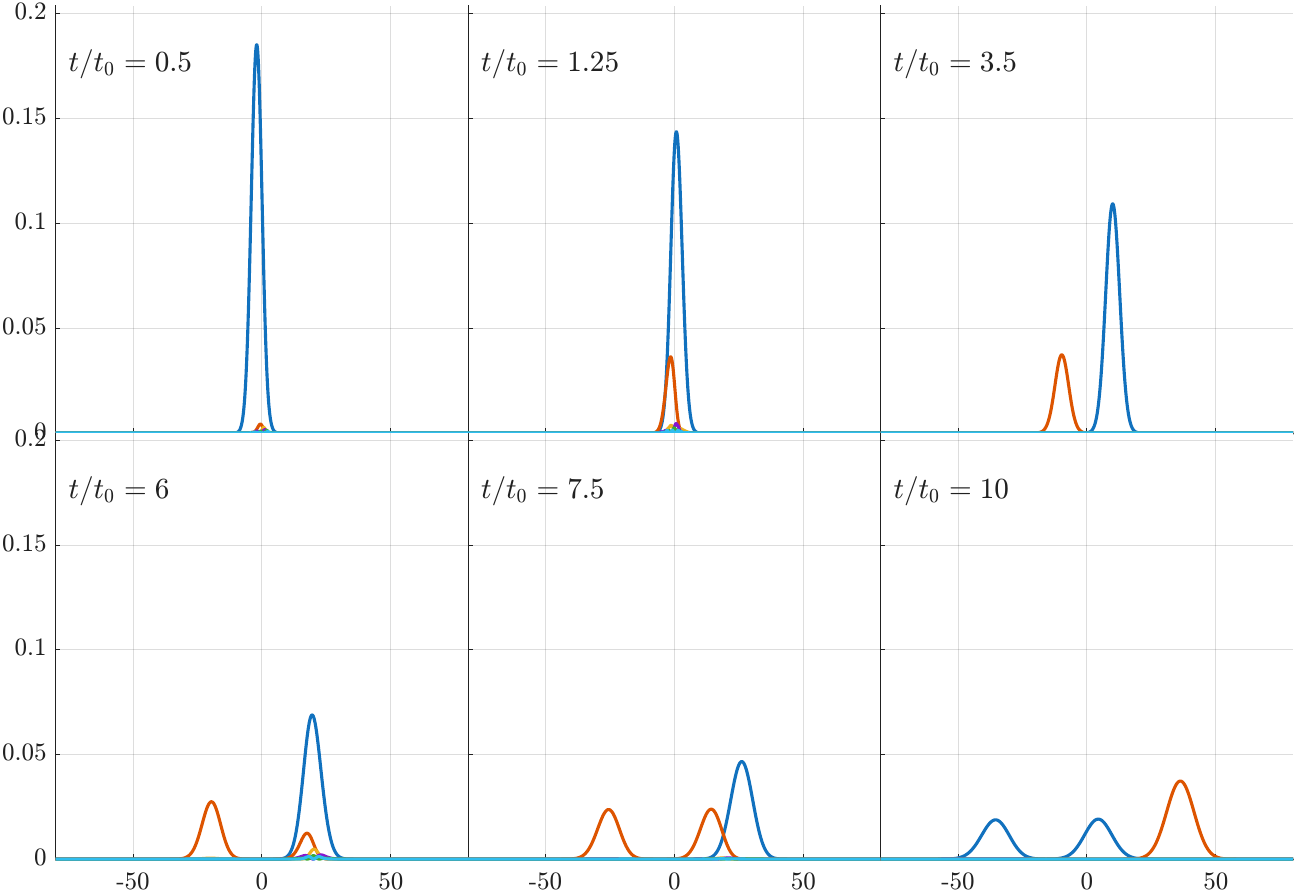}
    \caption{Evolution of $p_i(t)|\psi_i(x,t)|^2$ for the eigenvectors of the reduced density operator of the second particle (initially a single wavepacket) for the same scattering scenario.}
    \label{fig:double_peak_snapshots_p2}
\end{figure}

The corresponding evolution for the second particle is shown in Figure \ref{fig:double_peak_snapshots_p2}. Initially a single Gaussian, it splits into distinct wavepackets due to the sequential interactions with the two spatially separated components of the first particle. Thus, in the situation where scattering happens, the probe particle carries away the detailed information about the initial superposition state of the other particle. We will understand this analytically in the next section.

\section{Analytical analysis for narrow momentum spreads}
\label{sec:analytical_proof_general}

In the numerical investigations of one-dimensional scattering described above, the late-time entanglement spectrum is dominated by two eigenvalues adding nearly to one. Additionally, the wavefunction shape for the eigenstate corresponding to the reflected wavepacket appears very similar to that of the initial state of the environmental particle. In this section, we demonstrate analytically that these features are generic for any short-range potential for equal-mass particles, provided the momentum spread is narrow.

We consider two distinguishable particles of equal mass $m_1 = m_2 = m$ interacting via an arbitrary short-range potential $V(|x_1 - x_2|)$. Transforming to center-of-mass $X = (x_1 + x_2)/2$ and relative coordinates $x = x_1 - x_2$, the total Hamiltonian decouples as $H = H_{\text{CM}} + H_{\text{rel}}$. The center-of-mass component, $H_{\text{CM}} = -\frac{\hbar^2}{4m}\frac{\partial^2}{\partial X^2}$, describes free-particle propagation. The scattering dynamics are governed entirely by the relative Hamiltonian:
\begin{equation}
    H_{\text{rel}} = -\frac{\hbar^2}{2\mu}\frac{\partial^2}{\partial x^2} + V(|x|)
\end{equation}
where $\mu = m/2$ is the reduced mass. For an incident state with relative physical momentum $p$, solving this 1D scattering problem yields asymptotic relative transmission and reflection amplitudes, $t(p)$ and $r(p)$, which depend on the specific shape of the potential.

As in our numerical setup, we take the initial state of the full system to be a product state of two well-separated wavepackets. In momentum space, this is given by $\Psi_{\text{in}}(p_1, p_2) = \tilde{\psi}_1(p_1)\tilde{\psi}_2(p_2)$. While the numerical examples focused on symmetric collisions, this derivation holds for arbitrary initial central momenta $p_{c1}$ and $p_{c2}$. We assume that the momentum width $\Delta p$ is much smaller than the relative momentum $p_c = |p_{c1} - p_{c2}|/2$, and small enough that the variation of the scattering amplitudes $t(p)$ and $r(p)$ over the range $\Delta p$ around the central momentum is negligible. 

For equal-mass one-dimensional scattering, energy and momentum conservation constrain the asymptotic outgoing momenta to be either identical to the incoming momenta (transmission) or perfectly exchanged (reflection). The asymptotic outgoing state in momentum space is then approximately
\begin{equation}
    \Psi_{\text{out}}(p_1, p_2) \approx t(p_c)\tilde{\psi}_1(p_1)\tilde{\psi}_2(p_2) + r(p_c)\tilde{\psi}_1(p_2)\tilde{\psi}_2(p_1)
\end{equation}
where $p_c$ is the central value of the relative momentum $p_{\text{rel}} = (p_1 - p_2)/2$. Transforming back to position space, the late-time wavefunction is:
\begin{equation}
    \Psi(x_1, x_2, \tau) \approx t(p_c) \psi_1(x_1, \tau)\psi_2(x_2, \tau) + r(p_c) \psi_2(x_1, \tau)\psi_1(x_2, \tau)
\end{equation}
where $\tau \to \infty$ represents a time long after the collision. Here, $\psi_1(x, \tau)$ represents the freely propagated spatial profile originally associated with particle 1, and $\psi_2(x, \tau)$ represents the freely propagated spatial profile originally associated with particle 2.

The time-dependent density operator for the first particle can be computed by tracing out the spatial degrees of particle 2:
\begin{equation}
    \rho_1(x_1, x_1', \tau) = \int_{-\infty}^{\infty} \Psi(x_1, x_2, \tau) \Psi^*(x_1', x_2, \tau) \, dx_2
\end{equation}
Expanding this integral yields:
\begin{align}
    \rho_1(x_1, x_1', \tau) &\approx |t|^2 \psi_1(x_1, \tau)\psi_1^*(x_1', \tau) \int |\psi_2(x_2, \tau)|^2 \, dx_2 \nonumber \\
    &\quad + |r|^2 \psi_2(x_1, \tau)\psi_2^*(x_1', \tau) \int |\psi_1(x_2, \tau)|^2 \, dx_2 \nonumber \\
    &\quad + t r^* \psi_1(x_1, \tau)\psi_2^*(x_1', \tau) \int \psi_2(x_2, \tau)\psi_1^*(x_2, \tau) \, dx_2 \nonumber \\
    &\quad + t^* r \psi_2(x_1, \tau)\psi_1^*(x_1', \tau) \int \psi_1(x_2, \tau)\psi_2^*(x_2, \tau) \, dx_2
\end{align}

Assuming that the initial wavepackets are normalized  and have negligible overlap, the first two integrals evaluate to 1 and the last two integrals are negligible (since inner-products are preserved by time evolution). Thus, we have
\begin{equation}
    \rho_1(x_1, x_1', \tau) \approx |t|^2 \psi_1(x_1, \tau)\psi_1^*(x_1', \tau) + |r|^2 \psi_2(x_1, \tau)\psi_2^*(x_1', \tau)
\end{equation}
The expression on the right has exact eigenstates $\psi_1(x,t)$ with eigenvalue $|t|^2$ and $\psi_2(x,t)$ with eigenvalue $|r|^2$. Thus, to the extent that our approximations are valid, $\rho_1$ has exactly two eigenvalues adding to 1 (assuming both $t$ and $r$ are nonzero) and the shape of the wavepacket in the reflected branch is the same as the time-evolved initial wavepacket for the second particle.

\subsection{Extension to Two-Dimensional Scattering}

In this section, we provide some analytical insight into our numerical two-dimensional scattering results, making use of a toy calculation. 

In two dimensions, instead of simple reflection or transmission, we have a continuous angle $\theta$ parameterizing the outgoing direction of the scattered particles. For incoming particles with nearly opposite momenta, the outgoing momenta must also be nearly opposite by momentum conservation. It will be convenient to use an angular coordinate for the second particle that is shifted by $\pi$ from the angular coordinate of the first particle, so that $\theta_1$ and $\theta_2$ being equal corresponds to particles travelling in opposite directions.

As a crude model of the late-time wavefunction that enforces the momentum conservation, we can take
\[
\Psi(r_1,r_2, \theta_1, \theta_2) = \int d \theta S(\theta) R(r_1,t) \Theta(\theta_1 - \theta) R(r_2,t)  \Theta(\theta_2 - \theta) \; .
\]
Here,
\[
R(r_1,t) \Theta(\theta_1 - \theta)
\]
represents an outgoing wavepacket for the first particle centred at angle $\theta$. We are taking this to be a product of a radial wavefunction and an angular wavefunction as a simplifying approximation. The radial part is some function centred at $vt$ with a width that also increases linearly in $t$ (due to the uncertainty in momnetum of the initial particles). Because the outward travel and the wavepacket spreading should both be linear in $t$, the angular wavefunction $\Theta(\theta_1 - \theta)$ can be approximated as time-independent at late times. As a specific choice, we can take
\[
\Theta(\theta) \propto e^{- \theta^2 \over \delta^2}
\]
where we restrict to $\theta \in [-\pi , \pi]$. Our full model expression for the late-time wavefunction involves a product of this wavefunction with a corresponding one for the second particle with opposite central momentum, integrated over the scattering angle $\theta$ with a function $S(\theta)$ that captures the amplitude for scattering to different angles. This integration ensures that the late-time wavefunction is a quantum superposition of possible scattering outcomes. As a model example, we will take 
\[
S(\theta) = C e^{-{\theta^2 \over \Delta^2}}  \qquad \theta \in [-\pi, \pi] \; ,
\]
such that the scattering becomes more localized in the forward direction for smaller $\Delta$. 

Inserting our assumed explicit forms for the angular wavefunctions and the scattering amplitude function $S(\theta)$, we obtain
\[
\Psi(r_1,r_2,\theta_1, \theta_2) = R(r_1,t) R(r_2,t)   \int_{-\pi}^\pi d \theta e^{-{\theta^2 \over \Delta^2}}  e^{- (\theta_1 - \theta)^2 / \delta^2}  e^{- (\theta_2 - \theta)^2 / \delta^2} \; .
\]
Again as a simplifying assumption that will allow us to work with analytic expressions, we assume that $S(\theta)$ is localized enough (i.e. $\Delta$ is small enough) so that extending the range of integration for $\theta$ to the full real line is a good approximation. In this case, performing the integral gives
\[
\Psi(r_1,r_2,\theta_1, \theta_2) \propto  R(r_1,t) R(r_2,t)   \exp\left[
-\frac{\delta^2+\Delta^2}
{\delta^2(\delta^2+2\Delta^2)}
(\theta_1^2+\theta_2^2)
+
\frac{2\Delta^2}
{\delta^2(\delta^2+2\Delta^2)}
\theta_1 \theta_2
\right].
\label{eq:gaussianxy} \; .
\]
In order to understand the eigenvalues and eigenfunctions of the reduced density operator, we can perform a Schmidt decomposition of this expression using Mehler's formula. Defining normalized Hermite functions
\[
\varphi_n(x)
=
\left(\frac{\gamma}{\pi}\right)^{1/4}
\frac{1}{\sqrt{2^n n!}}\,
H_n(\sqrt{\gamma}\,x)\,
e^{-\gamma x^2/2},
\qquad n=0,1,2,\ldots \; ,
\]
Mehler's formula is
\[
\sum_{n=0}^{\infty}
t^n\,\varphi_n(x)\varphi_n(y)
=
\frac{\sqrt{\gamma/\pi}}{\sqrt{1-t^2}}
\exp\left[
-\frac{\gamma}{2}
\frac{1+t^2}{1-t^2}(x^2+y^2)
+
\frac{2\gamma t}{1-t^2}\,xy
\right],
\qquad |t|<1.
\]
Taking
\[
t = \frac{\sqrt{\delta^2+2\Delta^2}-\delta}
{\sqrt{\delta^2+2\Delta^2}+\delta} \qquad \qquad \gamma = \frac{2}
{\delta\sqrt{\delta^2+2\Delta^2}}
\]
we can apply this to our expression to give
\begin{equation}
\Psi(r_1,r_2,\theta_1, \theta_2) \propto  \sum_{n=0}^{\infty} \sqrt{\lambda_n} \,
[R(r_1,t) \varphi_n(\theta_1)][ R(r_2,t)\varphi_n(\theta_2)] \; .
\label{eq:schmidt}
\end{equation}
where we define
\[
\lambda_n=(1-t^2)t^{2n} \; .
\]
Since the $\varphi_n$ are orthogonal and the $\lambda_n$ sum to one, this expression gives a Schmidt decomposition of the original wavefunction. We can conclude that for the reduced density operator of particle $1$, the eigenstates are $u_n(r, \theta) = R(r_1,t) \varphi_n(\theta_1)$ with corresponding eigenvalue $\lambda_n$.
This analytical result reproduces several qualitative features of our numerical results. The $n$-th order Hermite function $H_n(x) e^{-x^2}$ has $n$ roots, so the probability density $|u_{n-1}|^2$ corresponding to the $n$th largest eigenvalue has $n$ angular peaks, as we observed in our numerics. Further, the peaks in the Hermite functions are arranged symmetrically about $\theta = 0$ with the largest peaks at the largest angles; this was also observed in our numerics.
Finally, the Hermite functions are increasingly spread out for increasing $n$, matching the increasing angular spread for decreasing eigenvalues that we saw in our numerics. 

\section{Discussion}

Studying the evolution of the eigenvalues and eigenvectors of the single-particle reduced density operator provides a fascinating perspective on quantum mechanical scattering. An initially pure wavepacket evolves into a discrete ensemble of late-time pure states, each with their own discrete probability and evolving wavefunction. This provides a detailed time-resolved perspective on what happens to a particle during scattering that goes well beyond the standard calculations of cross sections and angular distributions.

There are many possible directions for future work, including understanding quantitatively the effects on the eigenvalue spectrum and eigenstate evolution of varying the details of the wavepackets, the interaction potential, the mass ratio or (in two and higher dimensions) the impact parameter.

The scenario we have analyzed is the basic building block for many examples of  decoherence, where the interactions of a particle with its environment consist of a sequence of individual interactions with other particles. Each of these interactions should be similar to what we have studied, though may involve more general initial wavepackets that could be studied in future work. We have seen that an interaction between two localized particles sometimes produces an outgoing state that is a superposition of two or more wavepackets. We have seen that at least in a one dimensional example, the scattering of such a superposition state with another particle tends to re-localize the particle. It would be interesting to further explore this microscopic model of measurement/decoherence in higher dimensions. 

Through these future studies, it should be possible to understand in detail the typical evolution of the ensemble probabilities and eigenstates for a real-world particle in a dilute environment.  

\begin{acknowledgments}
This work is supported in part by the National Sciences and Engineering Research Council of Canada (NSERC) and the Simons Foundation via a Simons Investigator Award.
\end{acknowledgments}

\bibliographystyle{apsrev4-2}
\bibliography{references}

\end{document}